\documentclass[conference]{IEEEtran}
\usepackage{amsmath,amsfonts,amssymb,mathtools,dsfont}
\usepackage{graphicx,subfigure}
\usepackage{algorithmic,algorithm}
\usepackage{amsthm}
\usepackage[noadjust]{cite}

\newtheorem{lemma}{Lemma}

\newcommand{\Fopt}{{\mathbf{F}_\mathrm{opt}}}
\newcommand{\FRF}{{\mathbf{F}_\mathrm{RF}}}
\newcommand{\fopt}{{\mathbf{f}_\mathrm{opt}}}
\newcommand{\fRF}{{\mathbf{f}_\mathrm{RF}}}
\newcommand{\FBB}{{\mathbf{F}_\mathrm{BB}}}
\newcommand{\FBD}{{\mathbf{F}_\mathrm{BD}}}
\newcommand{\WBB}{{\mathbf{W}_\mathrm{BB}}}
\newcommand{\WRF}{{\mathbf{W}_\mathrm{RF}}}
\newcommand{\NRFr}{{N_\mathrm{RF}^\mathrm{r}}}
\newcommand{\NRFt}{{N_\mathrm{RF}^\mathrm{t}}}
\newcommand{\Nt}{{N_\mathrm{t}}}
\newcommand{\Nr}{{N_\mathrm{r}}}

\ifCLASSINFOpdf
\else
\fi
\IEEEoverridecommandlockouts
\begin{document}
\title{Alternating Minimization for Hybrid Precoding in Multiuser OFDM mmWave Systems}
\author{\IEEEauthorblockN{Xianghao Yu$^*$, Jun Zhang$^*$, and Khaled B. Letaief$^{*\dag}$, \emph{Fellow, IEEE} }
\IEEEauthorblockA{$^*$Dept. of ECE, The Hong Kong University of Science and Technology\\
	$^\dag$Hamad Bin Khalifa University, Doha, Qatar\\
Email: $^*$\{xyuam, eejzhang, eekhaled\}@ust.hk, $^\ddag$kletaief@hbku.edu.qa}
\thanks{This work was supported by the Hong Kong Research Grants Council under Grant No. 16210216. 
}}

\maketitle

\begin{abstract}
Hybrid precoding is a cost-effective approach to support directional transmissions for millimeter wave (mmWave) communications. While existing works on hybrid precoding mainly focus on single-user single-carrier transmission, in practice multicarrier transmission is needed to combat the much increased bandwidth, and multiuser MIMO can provide additional spatial multiplexing gains. In this paper, we propose a new hybrid precoding structure for multiuser OFDM mmWave systems, which greatly simplifies the hybrid precoder design and is able to approach the performance of the fully digital precoder. In particular, two groups of phase shifters are combined to map the signals from radio frequency (RF) chains to antennas. Then an effective hybrid precoding algorithm based on alternating minimization (AltMin) is proposed, which will alternately optimize the digital and analog precoders. A major algorithmic innovation is a LASSO formulation for the analog precoder, which yields computationally efficient algorithms. Simulation results will show the performance gain of the proposed algorithm. Moreover, it will reveal that canceling the interuser interference is critical in multiuser OFDM hybrid precoding systems.  
\end{abstract}


\IEEEpeerreviewmaketitle

\section{Introduction}
The spectrum crunch in current wireless systems stimulates extensive interests on exploiting new spectrum bands for cellular communications, and millimeter wave (mmWave) bands have been demonstrated to be a promising candidate in recent experiments \cite{6834753}.
Thanks to the smaller wavelength of mmWave signals, large-scale antenna arrays can be leveraged at both the transmitter and receiver sides, which can provide spatial multiplexing gains with the help of multiple-input multiple-output (MIMO) techniques.
On the other hand, the ten-fold increase of the carrier frequency introduces several challenges to mmWave communication systems, especially the high power consumption and cost of hardware components at mmWave bands \cite{rappaport2014millimeter}. In addition, the large available bandwidth at mmWave frequencies will result in wideband communication systems, where multicarrier techniques such as orthogonal frequency-division multiplexing (OFDM) will be utilized to overcome the frequency-selective fading.

By utilizing a small number of  radio frequency (RF) chains to combine a low-dimensional digital baseband precoder and another high-dimensional analog RF precoder, hybrid precoding stands out as a cost-effective transceiver solution \cite{6717211}. Moreover, to further reduce the power consumption in the RF domain, analog RF precoders are usually implemented by phase shifters, which induces a challenging unit modulus constraint and forms the major challenge in designing hybrid precoders. Given the large dimension of the design space and the unit modulus constraint, an important design aspect of hybrid precoders is the computational complexity.

There have been many recent works on hybrid precoder design in mmWave systems \cite{6717211,7397861,6884253,7448873,7037444,6928432,zhang2014achieving,7387790}. In \cite{6717211,7397861}, efficient hybrid precoding algorithms were developed for single-user single-carrier MIMO systems. The investigation was then extended to single-user OFDM \cite{7397861,6884253,7448873} and multiuser single-carrier systems \cite{7037444,6928432}. The main differences in these existing works are the approaches to deal with the unit modulus constraints of the analog precoder. Specifically, such constraints were tackled by orthogonal matching pursuit (OMP) in \cite{6717211,6884253,7037444}, by manifold optimization in \cite{7397861}, and by channel phase extraction in \cite{7448873,6928432}, respectively. There were also some studies on how to achieve the performance of the fully digital precoder with the hybrid structure \cite{zhang2014achieving,7387790}, yet requiring a large number of RF chains. Although various attempts have been made to balance the performance and computational complexity, there is no systematic approach to design computationally efficient hybrid precoders.

In this paper, we shall propose a novel hybrid precoder structure, which relaxes the unit modulus constraints of the analog part and thus significantly simplifies the hybrid precoder design. In particular, by adopting two groups of phase shifters to map the signals out of the RF chains to antennas, the constraints for the analog precoder become more tractable. We adopt the alternating minimization (AltMin) framework \cite{7397861} for the hybrid precoder design. While the digital part is similar to \cite{7397861}, the optimization of the analog RF precoder is formulated as a Least Absolute Shrinkage and Selection Operator (LASSO) problem, for which efficient algorithms are available. In addition, we discover that the hybrid precoder in the multiuser setting will produce residual interuser interference, as it only approximates the fully digital precoder. Such interference will significantly degrade the system performance, especially at high SNRs. This issue is more prominent in the multicarrier system as the analog precoder is shared by a large number of subcarriers. To this end, we propose to apply an additional block diagonalization (BD) precoder at the baseband to cancel the interuser interference, which is shown to be effective to further improve the spectral efficiency and multiplexing gain. Moreover, simulation results demonstrate that the proposed hybrid precoding algorithm can easily approach the performance of the fully digital precoder with a reasonable amount of RF chains.
\section{System Model and Problem Formulation}
\subsection{System Model}
Consider the downlink transmission of a multiuser OFDM mmWave MIMO system, where the base station (BS) is equipped with $N_\mathrm{t}$ antennas and transmits signals to $K$ $N_\mathrm{r}$-antenna users over $F$ subcarriers. On each subcarrier, $N_s$ data streams are transmitted to each user. The limitations of the RF chains are given by $KN_s\le N_\mathrm{RF}^\mathrm{t}\le N_\mathrm{t}$ and $N_s\le N_\mathrm{RF}^\mathrm{r}\le N_\mathrm{r}$, where $N_\mathrm{RF}^\mathrm{t}$ and $N_\mathrm{RF}^\mathrm{r}$ are the number of RF chains facilitated for the BS and each user, respectively.

The received signal for the $k$-th ($1\le k\le K$) user on the $f$-th subcarrier is given by
\begin{equation}
\mathbf{y}_{k,f}=\mathbf{W}^H_{\mathrm{BB},k,f}\mathbf{W}^H_{\mathrm{RF},k}\left(\mathbf{H}_{k,f}\sum_{i=1}^K\FRF\FBB_{k,f}\mathbf{s}_{k,f}+\mathbf{n}_{k,f}\right),
\end{equation}
where $\mathbf{s}_{k,f}\in\mathbb{C}^{N_s}$ is the transmitted symbol vector for the $k$-th user on the $f$-th subcarrier such that $\mathbb{E}[\mathbf{s}_{k,f}\mathbf{s}^H_{k,f}]=\frac{1}{KN_sF}\mathbf{I}_{N_s}$. The digital baseband precoders and combiners are symbolized by $\FBB_{k,f}\in\mathbb{C}^{\NRFt\times N_s}$ and $\WBB_{k,f}\in\mathbb{C}^{\NRFr\times N_s}$, respectively.
Because the transmitted signals for all the users are mixed together via the digital baseband precoder and the analog RF precoder is a post-IFFT (inverse fast Fourier transform) operation, the analog RF precoder is shared by all the users and subcarriers, denoted as $\FRF\in\mathbb{C}^{\Nt\times \NRFt}$. Similarly, the analog RF combiner is a subcarrier-independent operation for each user $k$, denoted as $\WRF_k\in\mathbb{C}^{\Nr\times \NRFr}$. Furthermore, the additive noise at the users is represented by $\mathbf{n}_{k,f}\in\mathbb{C}^{\Nr}$, whose elements are independent and identically distributed (i.i.d.) according to $\mathcal{CN}(0, \sigma^2)$. The mmWave MIMO channel between the BS and the $k$-th user on the $f$-th subcarrier, denoted as $\mathbf{H}_{k,f}$, can be characterized by the Saleh-Valenzuela model as \cite{6717211}
\begin{equation}
\mathbf{H}_{k,f}=\gamma_k\sum_{i=0}^{N_{\mathrm{cl},k}-1} \sum_{l=1}^{N_{\mathrm{ray},k}}{\alpha_{il,k}\mathbf{a}_r(\theta_{il,k})\mathbf{a}_t^H(\phi_{il,k})}e^{-j2\pi if/F}.
\end{equation}
The normalization factor $\gamma_k$ is specified by $\gamma_k=\sqrt{\frac{\rho_k\Nt\Nr}{N_{\mathrm{cl},k}N_{\mathrm{ray},k}}}$, where $N_{\mathrm{cl},k}$ and $N_{\mathrm{ray},k}$ represent the number of clusters and the number of rays in each cluster, and $\rho_k$ is the path loss between the BS to the $k$-th user.
The gain of the $l$-th ray in the $i$-th propagation cluster is denoted as $\alpha_{il,k}$. In addition, $\mathbf{a}_r(\theta_{il,k})$ and $\mathbf{a}_t(\phi_{il,k})$ stand for the receive and transmit array response vectors with the corresponding angle of arrival $\theta_{il,k}$ and angle of departure $\phi_{il,k}$. The detailed expressions can be found in \cite{6717211}.

\subsection{Analog RF Precoder Structure}
\begin{figure}[tbp]
	\centering
	\subfigure[Conventional analog RF precoder structure.]{
		\includegraphics[width=6cm]{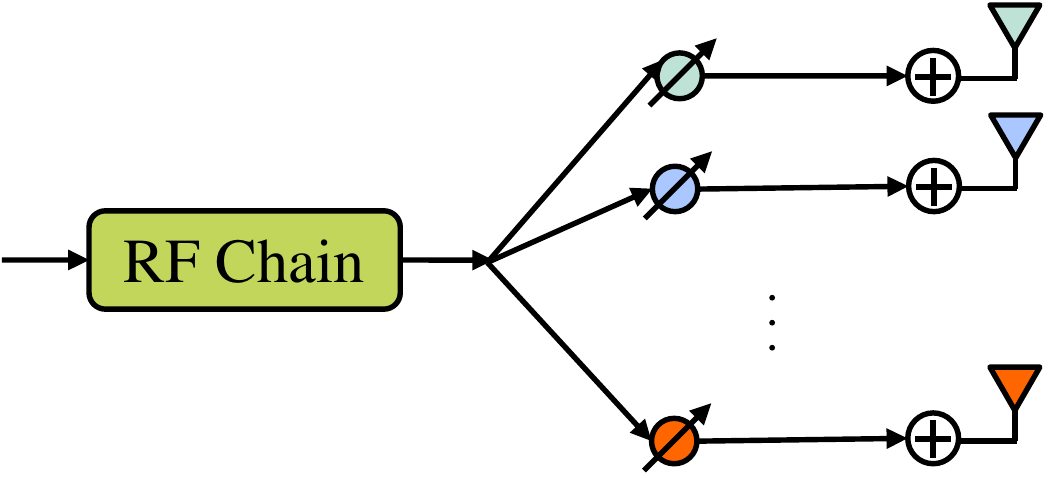}\label{fig11}
	}
	\subfigure[DPS analog RF precoder structure.]{
		\includegraphics[width=6cm]{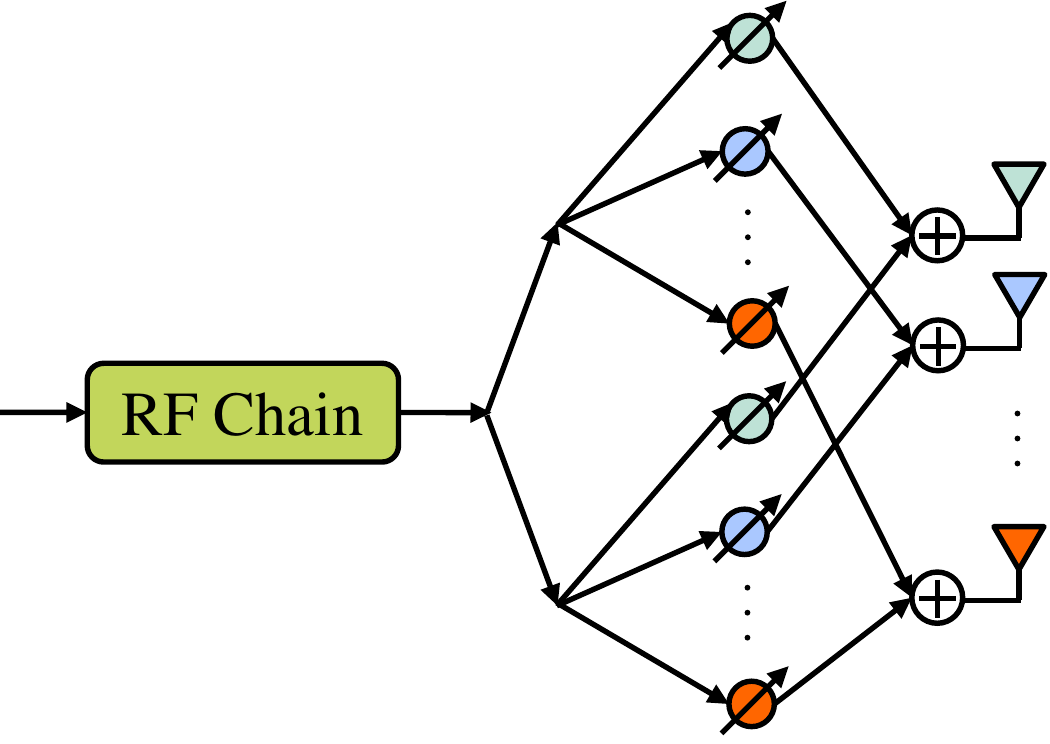}\label{fig12}
	}
	\caption{Comparison of two analog RF precoder structures.}
\end{figure}

As mentioned before, the analog precoder is practically implemented by phase shifters. Conventionally, in either the fully- or partially-connected structure \cite{7397861}, each route from a certain RF chain to one connected antenna element is implemented by a phase shifter, as shown in Fig. \ref{fig11}. This mapping strategy implies that each nonzero element in the analog precoding and combining matrices should have unit modulus, i.e., $|(\FRF)_{i,j}|=|(\WRF)_{i,j}|=1$. This is intrinsically a non-convex constraint and difficult to tackle with, which forms the main design challenge.

In this paper, we propose a new analog RF precoder structure, as shown in Fig. \ref{fig12}, where the phase shifter network is divided into two groups, referred as the \emph{Double Phase Shifter (DPS) structure}. For each route from an RF chain to an antenna element, a unique phase shifter in each group will be selected and summed up together to compose the analog precoding gain. Under this special structure, each nonzero element in the analog RF precoding matrix corresponds to a sum of two phase shifters. In other words, the new constraints for the analog RF precoder and combiner are $|(\FRF)_{i,j}|\le 2$ and $|(\WRF)_{i,j}|\le 2$ since the amplitude of a sum of two phase shifters should be less than 2. By doubling the number of phase shifters, the new constraint becomes convex and therefore makes it more tractable and promising to develop low-complexity design approaches.

\subsection{Problem Formulation}
As shown in \cite{6717211,7397861}, minimizing the Euclidean distance between the fully digital precoder and the hybrid precoder is an effective way to design the hybrid precoder in mmWave MIMO systems, whose formulation\footnote{Here we focus on the precoder design, and the combiner is designed in the same way without the transmit power constraint.} is given by
\begin{equation}\label{problemformulation}
\begin{aligned}
&\underset{\mathbf{F}_\mathrm{RF},\mathbf{F}_\mathrm{BB}}{\mathrm{minimize}} && \left\Vert \mathbf{F}_\mathrm{opt}-\mathbf{F}_\mathrm{RF}\mathbf{F}_\mathrm{BB}\right\Vert _F\\
&\mathrm{subject\thinspace to}&&
\begin{cases}
|(\FRF)_{i,j}|\le 2\\
\left\|\mathbf{F}_\mathrm{RF}\mathbf{F}_\mathrm{BB}\right\|_F^2\le KN_sF,
\end{cases}
\end{aligned}
\end{equation}
where $\Fopt=\left[\Fopt_{1,1},\cdots,\Fopt_{k,f},\cdots,\Fopt_{K,F}\right]$ is the combined fully digital precoder with dimension $\Nt\times KN_sF$, and $\FBB=\left[\FBB_{1,1},\cdots,\FBB_{k,f},\cdots,\FBB_{K,F}\right]$ is the concatenated digital precoder with dimension $\NRFt\times KN_sF$. The second constraint is the transmit power constraint at the BS side.

Problem \eqref{problemformulation} is a matrix decomposition problem, and the goal of this formulation is to find an accurate approximation for an arbitrary fully digital precoder. With this formulation, the proposed algorithm can be applied with any fully digital precoder. It also has been shown in \cite{6717211} that minimizing the objective function
in \eqref{problemformulation} leads to the maximization of spectral efficiency in the single-user single-carrier case. In this paper, we will adopt the classical BD precoder as $\Fopt$, which is asymptotically optimal in the high signal-to-noise ratio (SNR) regime \cite{1261332}.

\section{Hybrid Precoder Design}
Alternating minimization, which separates the optimization of the objective function with respect to different variable subsets in each step, has been shown to be effective in hybrid precoding \cite{7397861} and various other applications, e.g., matrix completion, phase retrieval, and dictionary learning \cite{7130654}. In this section, we will adopt alternating minimization as the main approach to design the hybrid precoder under the DPS analog RF precoder structure.

\subsection{Single-carrier Systems}
While the main focus of this paper is on multiuser multicarrier systems, some advantages of the proposed DPS structure will be firstly presented in single-carrier systems, as shown in the following result.
\begin{lemma}\label{lem1}
	For single-carrier systems, with the DPS precoder structure in Fig. \ref{fig12}, the fully digital precoder $\Fopt$ can be perfectly decomposed into the hybrid precoder $\FRF$ and $\FBB$ using the minimum number of RF chains, i.e., $\NRFt=KN_s$ and $\NRFr=N_s$. 
\end{lemma}
\begin{IEEEproof}
	The proof can be easily obtained by the rank sufficiency of $\FRF$ and $\FBB$ in the decomposition when $F=1$, and is omitted due to space limitation.
\end{IEEEproof}

Lemma \ref{lem1} shows that, for single-carrier systems with either single-user or multiuser transmissions, the performance of the fully digital precoder can be easily obtained with a hybrid precoder via a simple matrix decomposition. Note that, with the conventional analog precoder structure, the number of RF chains should be at least twice that of the data streams in order to achieve the fully digital precoder, i.e., $\NRFt=2KN_s$ and $\NRFr=2N_s$ \cite{zhang2014achieving,7397861}. Considering that the RF chain is significantly more power hungry than the phase shifter \cite{7397861}, the proposed structure is more energy efficient when achieving the fully digital precoder.

\subsection{Hybrid Precoder Design in Multicarrier Systems via Alternating Minimization}
In each step of alternating minimization, one part of the hybrid precoder is fixed while the other part will be optimized.  Since the main difficulty is the constraint on the analog precoder, we will focus on the analog precoder design in the following. The optimization of the analog precoder is given by
\begin{equation}\label{analogp}
\begin{aligned}
&\underset{\mathbf{F}_\mathrm{RF}}{\mathrm{minimize}} && \left\Vert \mathbf{F}_\mathrm{opt}-\mathbf{F}_\mathrm{RF}\mathbf{F}_\mathrm{BB}\right\Vert _F\\
&\mathrm{subject\thinspace to}&&
|(\FRF)_{i,j}|\le 2.
\end{aligned}
\end{equation}
Note that the power constraint in \eqref{problemformulation} is temporarily removed. In fact, a simple normalization operation can be adopted if the power constraint is not satisfied, which will be discussed later.
The optimization problem \eqref{analogp} is a convex problem and can be solved by solvers such as CVX. Nevertheless, to further reduce the computational complexity, we will illustrate the inherent structure of the solution by considering the dual problem. This will lead to a closed-form solution to \eqref{analogp}.
\begin{lemma}\label{lem2}
	The dual problem of \eqref{analogp} is a LASSO problem, given by
	\begin{equation}\label{lasso}
	\underset{\mathbf{x}}{\mathrm{minimize}} \quad \frac{1}{2}\left\Vert 
	\mathbf{Ax-b}\right\Vert _2^2+2\Vert\mathbf{x}\Vert_1.
	\end{equation}
	The parameters $\mathbf{A}$ and $\mathbf{b}$ are given by
	\begin{equation}
	\mathbf{A}=\mathbf{S}^{\frac{1}{2}}\mathbf{U},\quad\mathbf{b}=\mathbf{AD}^H\fopt,
	\end{equation}
	where $\mathbf{D}=\mathbf{F}_\mathrm{BB}^T\otimes \mathbf{I}_{\Nt}$ and $\left(\mathbf{D}^H\mathbf{D}\right)^{-1}=\mathbf{USU}^H$ is the singular value decomposition (SVD) of $\left(\mathbf{D}^H\mathbf{D}\right)^{-1}$. The optimal solution of \eqref{analogp} can be written by
	\begin{equation}
	\mathbf{f}_\mathrm{RF}^\star=\mathbf{A}^H\left(\mathbf{b}-\mathbf{Ax}^\star\right),
	\end{equation}
	where $\fRF=\mathrm{vec}(\FRF)$.
\end{lemma}
\begin{IEEEproof}
The proof is omitted due to space limitation.
\end{IEEEproof}
Based on Lemma \ref{lem2}, the analog precoder design problem is transferred to a LASSO problem. This provides the opportunity to leverage the large body of existing works on efficiently solving the general LASSO problem \cite{hastie2015statistical}. Here we are interested in a special case where we can get a closed-form solution to the problem, which will significantly reduce the computational complexity of the hybrid precoding algorithm.

It was shown in \cite{7397861} that enforcing a semi-orthogonal constraint to the digital precoder will incur little performance loss in single-user multicarrier systems. Inspired by this work, we resort to a similar approach, i.e., imposing a semi-orthogonal constraint to the digital precoder, which is specified as
\begin{equation}\label{semio}
\FBB\mathbf{F}_\mathrm{BB}^H=\mathbf{I}_\NRFt.
\end{equation}
Under this constraint, the observation matrix $\mathbf{A}$ in the LASSO problem \eqref{lasso} is also semi-orthogonal, i.e.,
\begin{equation}
\begin{split}
\mathbf{A}^H\mathbf{A}&=\left(\mathbf{D}^H\mathbf{D}\right)^{-1}\\
&=\left((\mathbf{F}_\mathrm{BB}^T\otimes\mathbf{I}_{\NRFt})^H(\mathbf{F}_\mathrm{BB}^T\otimes\mathbf{I}_{\NRFt})\right)^{-1}\\
&=\left((\mathbf{F}_\mathrm{BB}^*\otimes\mathbf{I}_{\NRFt})(\mathbf{F}_\mathrm{BB}^T\otimes\mathbf{I}_{\NRFt})\right)^{-1}\\
&=\left((\FBB\mathbf{F}_\mathrm{BB}^H)^T\otimes\mathbf{I}_{\NRFt}\right)^{-1}=\mathbf{I}_{\NRFt^2}.
\end{split}
\end{equation}
With the semi-orthogonal observation matrix $\mathbf{A}$, the LASSO problem \eqref{lasso} has a closed-form solution as \cite{hastie2015statistical}
\begin{equation}
\mathbf{x}^\star=\exp\{j\angle(\mathbf{A}^H\mathbf{b})\}\left(\left|\mathbf{A}^H\mathbf{b}\right|-2\right)^+,
\end{equation}
where $(x)^+=\max\{0,x\}$, and the corresponding solution to $\FRF$ in \eqref{analogp} is
\begin{equation}\label{lassoso}
\mathbf{F}_\mathrm{RF}^\star=\Fopt\mathbf{F}_\mathrm{BB}^H-\exp\left\{j\angle\left(\Fopt\mathbf{F}_\mathrm{BB}^H\right)\right\}\left(\left|\Fopt\mathbf{F}_\mathrm{BB}^H\right|-2\right)^+.
\end{equation}
Note that, to obtain an analog precoder $\FRF$ with the fixed digital precoder $\FBB$, a product between $\Fopt$ and $\mathbf{F}_\mathrm{BB}^H$ is the only required step, which is much more computationally efficient than solving the original problem \eqref{analogp} using an algorithm-embedded solver.  

After updating the analog precoder $\FRF$, optimizing the digital precoder $\FBB$ with the semi-orthogonal constraint \eqref{semio} is a typical semi-orthogonal Procrustes problem (OPP). The solution is similar to \cite[Eq. 28]{7397861}, which can be expressed as
\begin{equation}
\FBB = \mathbf{VU}_1^H,\label{OPP}
\end{equation}
where $\mathbf{U}_1\mathbf{SV}^H=\mathbf{F}_\mathrm{opt}^H\FRF$ is the SVD of $\mathbf{F}_\mathrm{opt}^H\FRF$, and $\mathbf{S}$ is a diagonal matrix whose nonzero elements are the first $\NRFt$ nonzero singular values $\sigma_1,\cdots,\sigma_{\NRFt}$.

\subsection{Interuser Interference Cancellation}
While we can perfectly cancel the interuser interference with the fully digital precoder, there will be residual interuser interference for the hybrid precoder, which is an approximation of the fully digital one. Later in Section \ref{IV}, we will see that in multiuser multicarrier systems, interuser interference is a severe problem that will dramatically degrade the hybrid precoding performance, especially at high SNRs.

In this subsection, after designing the hybrid precoder and combiner, we propose to cascade another digital baseband precoder $\FBD$ that is responsible for canceling the residual interuser interference. In particular, with the hybrid precoder and combiner at hand, we define an effective channel for the $k$-th user on the $f$-th subcarrier as
\begin{equation}\label{effch}
\mathbf{\hat H}_{k,f}={\mathbf{W}^H_\mathrm{BB}}_{k,f}{\mathbf{W}^H_\mathrm{RF}}_{k}\mathbf{H}_{k,f}{\FRF}{\FBB}_{f},
\end{equation}
where $\FBB_f=\left[\FBB_{1,1},\cdots,\FBB_{k,f},\cdots,\FBB_{K,f}\right]\in\mathbb{C}^{\NRFt\times KN_s}$ is the composite digital precoder on the $f$-th subcarrier, and $\mathbf{\hat H}_{k,f}\in\mathbb{C}^{N_s\times KN_s}$ is the effective channel. Our goal is to design the precoders $\FBD_{k,f}$, which satisfy the conditions
\begin{equation}
\mathbf{\hat H}_{j,f}\FBD_{k,f}=\mathbf{0}, \quad k\ne j.
\end{equation}
A simple way to achieve the conditions is the BD precoder, and note that the dimension of the effective channel is sufficient to perform the BD design. More details can be found in \cite{1261332}.

The overall hybrid precoding algorithm is summarized as the LASSO-AltMin algorithm.
\floatname{algorithm}{LASSO-AltMin Algorithm:}
\begin{algorithm}[h]
	\caption{LASSO Based Alternating Minimization Algorithm}
	\label{alternating}
	\begin{algorithmic}[1]
		\REQUIRE
		$\mathbf{F}_{\mathrm{opt}}$
		\STATE Construct a feasible $\mathbf{F}_\mathrm{RF}^{(0)}$ and set $k=0$;
		\REPEAT 
		\STATE Fix $\FRF^{(k)}$, solving $\FBB^{(k)}$ using the solution to OPP \eqref{OPP};
		\STATE Fix $\FBB^{(k)}$, and update $\FRF^{(k+1)}$ by the LASSO solution \eqref{lassoso};
		\STATE $k\leftarrow k+1$;
		\UNTIL a stopping criterion triggers.

\STATE Compose the effective channels $\mathbf{\hat H}_{k,f}$ according to \eqref{effch}.
\STATE Compute BD precoders $\FBD_{k,f}$ \cite{1261332}.
\STATE The final digital baseband precoder ${{\mathbf{F}}_\mathrm{B}}_{k,f}=\FBB_{k,f}\FBD_{k,f}$.
\STATE  For the digital precoder at the transmit end, normalize
$\widehat{\mathbf{F}}_\mathrm{B}=\frac{\sqrt{KN_sF}}{\left\Vert\mathbf{F}_\mathrm{RF}\mathbf{F}_\mathrm{B}\right\Vert_F}\mathbf{F}_\mathrm{B}$ if the power constraint in \eqref{problemformulation} is not satisfied.
	\end{algorithmic}
\end{algorithm}
Note that Steps 3 and 4 both give the globally optimal solution to the digital and analog precoders, respectively. Hence, the algorithm will converge to a stationary point of problem \eqref{analogp} with an additional constraint \eqref{semio}, since it is a two block coordinate descent procedure \cite{grippo2000convergence}. In the last step, we normalize the digital precoder if the transmit power constraint is not satisfied. It has been shown in \cite[Lemma 1]{7397861} that as long as we can make the Euclidean distance between the optimal digital precoder and the hybrid precoder sufficiently small when ignoring the power constraint, the normalization step will also achieve a small distance to the optimal digital precoder.

\section{Simulation Results}\label{IV}
In this section, we will evaluate the performance of the proposed LASSO-AltMin algorithm and compare it with the OMP algorithm \cite{6717211,7037444} in multiuser OFDM mmWave systems. Assume that $N_s=3$ data streams are sent from the BS to each user in a 3-user ($K=3$) MIMO systems over $F=128$ subcarriers, with $\Nt=256$ and $\Nr=16$, while both are equipped with uniform linear arrays (ULAs). The channel parameters are given by $N_{\mathrm{cl},k}=3$ clusters and $N_{\mathrm{ray},k}=8$ rays.
The angles of departure and arrival (AoDs and AoAs) follow the Laplacian distribution with uniformly distributed mean angles in $[0,2\pi]$ and angular spread of 10 degrees. The antenna elements in ULAs are separated by a half wavelength distance, and all simulation results are averaged over 5000 channel realizations.
\begin{figure}[htbp]
	\centering
	\includegraphics[width=7cm]{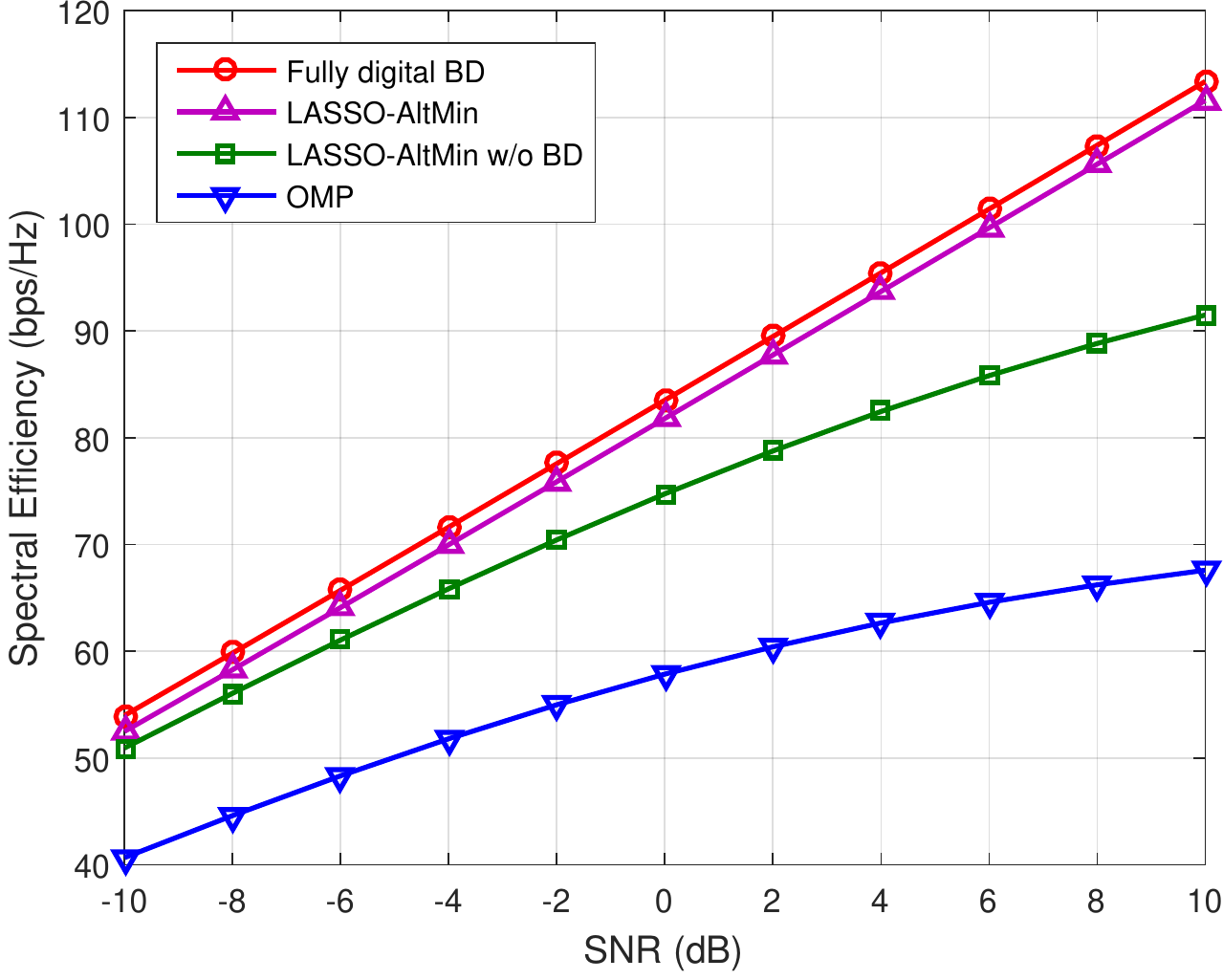}
	\caption{Spectral efficiency achieved by different precoding algorithms when $\NRFt=K\NRFr=KN_s$.}\label{fig1}
\end{figure}

Fig. \ref{fig1} shows the spectral efficiency of different algorithms with the minimum numbers of RF chains, i.e., $\NRFt=KN_s$ and $\NRFr=N_s$. First, we see that the proposed algorithm significantly outperforms the existing OMP algorithm implemented under the conventional analog precoder structure. The performance gain mainly comes from doubling the number of phase shifters, which provides more degrees of freedom for the analog precoding gain. We observe in the simulation that the time complexity of the proposed LASSO-AltMin algorithm is comparable with the OMP algorithm, mainly thanks to the closed-form solutions in the alternating procedures.

In \cite{7397861,7037444}, it has been pointed out that approximating the fully digital precoder with a hybrid structure will lead to a near optimal performance in single-user single-carrier, single-user multicarrier, and multiuser single-carrier mmWave MIMO systems. 
In Fig. \ref{fig1}, we evaluate the performance of the LASSO-AltMin algorithm without the additional BD operation. We discover that, without the BD precoder canceling the interuser interference, there will be residual interuser interference, which results in an obvious performance loss compared to the fully digital one, especially at high SNRs. This phenomenon illustrates that simply approximating the fully digital precoder with the hybrid one is not sufficient in multiuser multicarrier mmWave systems since the analog precoder is shared by a large number of subcarriers. The comparison in Fig. \ref{fig1} demonstrates the effectiveness and necessity of the BD steps in the proposed LASSO-AltMin algorithm.

\begin{figure}[htbp]
	\centering
	\includegraphics[width=7cm]{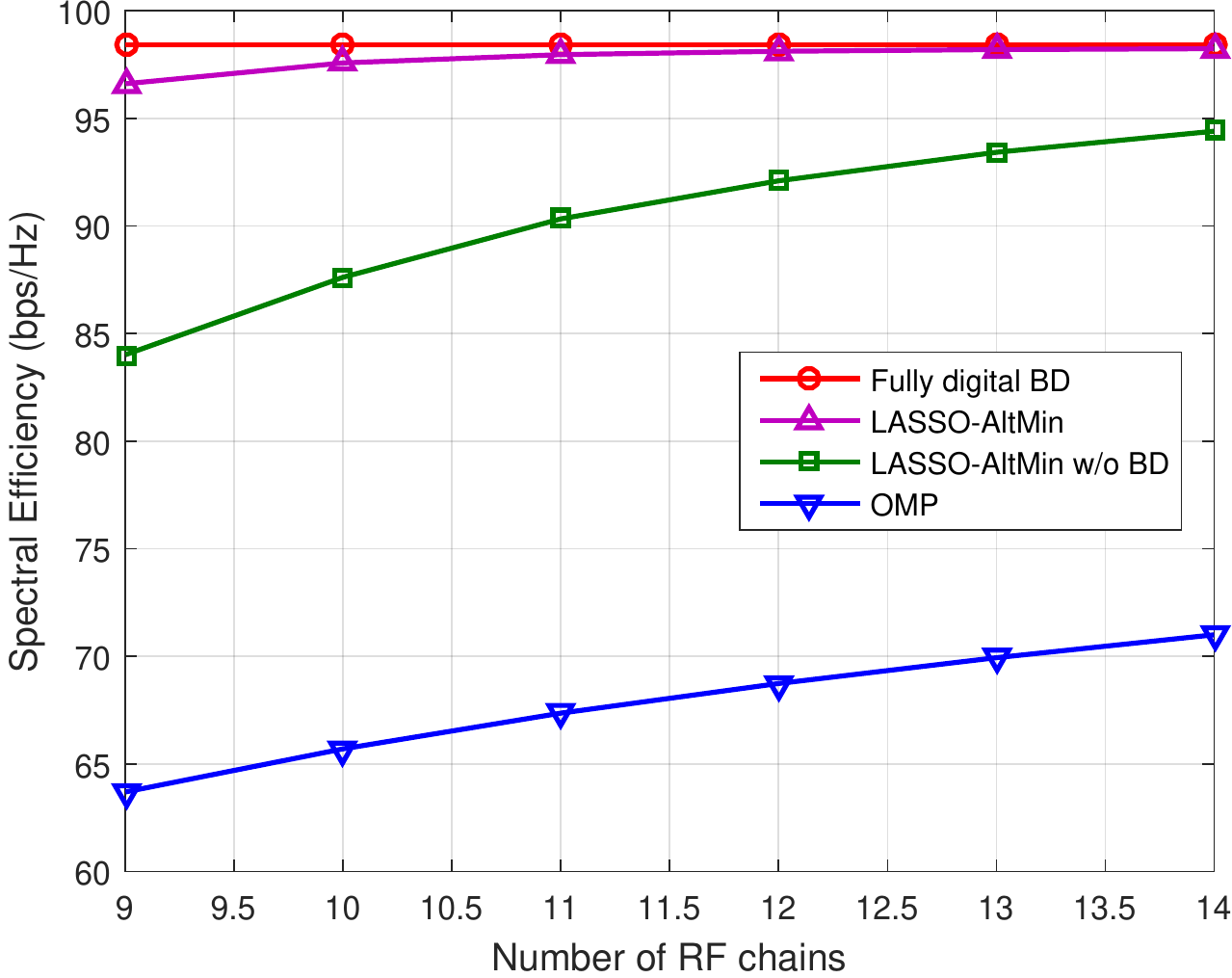}
	\caption{Spectral efficiency achieved by different precoding algorithms for different transmit RF chain numbers $\NRFt$, given $\NRFr=N_s$ and SNR=5 dB.}\label{fig2}
\end{figure}

Fig. \ref{fig2} compares the performance of different precoding schemes for different RF chain numbers $\NRFt$ at the BS side while keeping $\NRFr=N_s$ as the minimum number of RF chains at each user.
It is shown that the proposed LASSO-AltMin algorithm can approach the performance of the fully digital precoder when the number of RF chains is slightly larger than the number of data streams, which cannot be realized by the existing OMP algorithm. Thanks to the newly proposed DPS analog precoder structure and the LASSO-AltMin algorithm, it turns out that there will not be much performance loss when we adopt hybrid precoding in multiuser OFDM mmWave systems.

\section{Conclusions}
This paper proposed a new analog precoder structure for hybrid precoding, based on which a LASSO based alternating minimization algorithm was proposed for hybrid precoder design in multiuser OFDM systems. The paper helped unravel some valuable design insights:
\begin{itemize}
	\item It is beneficial, from both performance and complexity points of view, to implement twice the number of phase shifters in the analog precoder, as shown in Fig. \ref{fig12}. 
	\item Different from other hybrid precoding systems, in multiuser multicarrier systems, interuser interference is a vital problem that we should deal with in addition to the fully digital precoder approximation. To solve this problem, it is effective to cascade a digital baseband precoder that specializes in canceling the interuser interference.
\end{itemize}
It is interesting to extend the proposed DPS analog precoder structure to investigate other problems involving hybrid precoder design, e.g., to consider the hybrid precoder design combined with channel training and feedback.


\bibliographystyle{IEEEtran}
\bibliography{bare_conf}
\end{document}